\documentclass[10pt,letterpaper,twocolumn,aps,prd,groupedaddress,notitlepage,nofootinbib,showkeys]{revtex4-1}

\pdfoutput=1

\usepackage{color}
\usepackage{array}
\usepackage{prettyref}
\usepackage{float}
\usepackage{mathtools}
\usepackage{enumitem}
\usepackage{multirow}
\usepackage{amsmath}
\usepackage{amssymb}

\usepackage[unicode=true,bookmarks=false,pdfborder={0 0 0},
pdfborderstyle={},backref=false,colorlinks=true,
citecolor=darkBlue,linkcolor=darkBlue,urlcolor=darkBlue
]{hyperref}

\makeatletter

\pdfpageheight\paperheight
\pdfpagewidth\paperwidth

\providecommand{\tabularnewline}{\\}

\usepackage{amsfonts}
\definecolor{darkBlue}{rgb}{0,0,0.75}

\usepackage{tensind} 
\tensordelimiter{?}

\allowdisplaybreaks

\makeatother

\begin{document}

\title{Algebraic Properties of Einstein Solutions in Ghost-Free Bimetric
Theory}

\author{Mikica Kocic}

\author{Marcus H\"{o}g\r{a}s}

\author{Francesco Torsello}

\author{Edvard M\"{o}rtsell}

\affiliation{Department of Physics \& The Oskar Klein Centre, Stockholm University,
AlbaNova University Centre, SE-106 91 Stockholm, Sweden}
\begin{abstract}
A known fact is that an Einstein solution in one sector in ghost-free
bimetric theory implies an Einstein solution in the other sector.
Earlier studies have also shown that some classes of bimetric models
necessitate proportional solutions between the sectors. Here we consider
a general setup of the parameters in the theory as well as the general
algebraic form of the potential. We show that, if one sector has an
Einstein solution, the solutions are either proportional or block
proportional with at most two different eigenvalues in the square
root governing metric interactions.
\end{abstract}
\maketitle

\newcommand{\bSe}{\begin{subequations}} 
\newcommand{\eSe}{\end{subequations}}

\global\long\def\dd{\mathrm{d}}
\global\long\def\ee{\mathrm{e}}
\global\long\def\ii{\mathrm{i}}
\global\long\def\diag{\operatorname{diag}}
\global\long\def\eH{\mathrm{{\scriptscriptstyle H}}}
\global\long\def\tudu#1#2#3#4{?{\mbox{\ensuremath{#1}}}^{#2}{}_{#3}{}^{#4}?}
\global\long\def\tdud#1#2#3#4{?{\mbox{\ensuremath{#1}}}{}_{#2}{}^{#3}{}_{#4}?}
\global\long\def\tud#1#2#3{?{\mbox{\ensuremath{#1}}}^{#2}{}_{#3}?}
\global\long\def\tdu#1#2#3{?{\mbox{\ensuremath{#1}}}{}_{#2}{}^{#3}?}

\section{Introduction and Summary\vspace{-0.5em}}

The understanding of classical interacting massive spin-2 theories
has seen a significant progress in recent years. Several consistent
theories have emerged, free of the instability known as the Boulware-Deser
ghost \cite{Boulware:1973my}. In particular, de Rham-Gabadadze-Tolley
(dRGT) massive gravity was introduced in \cite{deRham:2010ik,deRham:2010kj}
as a nonlinear theory of a massive spin-2 field (shown to be ghost-free
in \cite{Hassan:2011hr}), and the Hassan-Rosen (HR) bimetric theory
was introduced in \cite{Hassan:2011zd,Hassan:2011ea} as a ghost-free
nonlinear theory of two interacting spin-2 fields. Recent reviews
of these theories and their extensions (like vielbein formulation,
multiple interacting spin-2 fields etc.), can be found in \cite{Schmidt-May:2015vnx,deRham:2014zqa}.

In this paper we focus on the HR bimetric theory, which schematically
consists of two GR sectors governed by their own metric, here denoted
$g$ and $f$, coupled through a ghost-free bimetric interaction term
that involves the square root of the two metrics, $S=\sqrt{g^{-1}f}$.
In addition, the interaction term is parameterized by real constants,
$\beta_{n}$, where $n=0,...,4$ in four dimensions (more details
are in \prettyref{sec:background}). 

As shown in \cite{Hassan:2014vja}, an Einstein solution in one sector
of the HR bimetric theory is a necessary condition to have an Einstein
solution in the other sector. This bi-Einstein setup, if attainable,
imposes certain algebraic conditions on the bimetric interaction term
and limits the structure of the square root matrix depending on the
$\beta$-parameters. In \cite{Hassan:2014vja}, it was also shown
that the two metrics are proportional for some classes of bimetric
models.

Assuming a general form of the square root, we provide the following
statement.\vspace{0.7em}

\noindent \textbf{Proposition}. \textit{In four dimensions, unless
algebraically decoupled, Einstein solutions to the bimetric equations
are either proportional or block proportional with at most two different
eigenvalues in the square root.}\vspace{0.7em}

\noindent The proof is given in \prettyref{sec:proof}. The algebraically
decoupled cases are handled in \prettyref{sec:decoupled}.

The proof is based on the algebraic classification of the square roots
of $g^{-1}f$ given in \cite{Hassan:2017a}. By this classification,
we get: for Type I (a diagonal square root) either one or two eigenvalues,
for Type IIa (Jordan block of size 2) a single eigenvalue but with
the constrained $\beta$-parameters, for Type IIb (a complex block)
it either falls back to Type I or has no solutions, for Type III (Jordan
block of size 3) a single eigenvalue but with the constrained $\beta$-parameters.
Type IV (a non-primary square root) has the same structure as Type
IIb. 

A comprehensive list of all possible square roots is summarized in
Table \ref{tab:segre} in terms of Segre characterization.\footnote{The Segre characteristic is a set of integers listed in descending
order that give the sizes of the blocks in a Jordan normal form. Complex
Jordan blocks are denoted by $z\bar{z}$ instead. The integers corresponding
to submatrices containing the same eigenvalue are grouped together
in parentheses. For example, $[(21)1]$ is a class of matrices which
have two different eigenvalues, where the first is in a sequence of
Jordan blocks of sizes 2 and 1, and the second is in a Jordan block
of size 1.}

\begin{table}[H]
\noindent \centering{}\bgroup\def\arraystretch{1.5}%
\begin{tabular}{|c|c|c|c|c|}
\hline 
\,Type\, & \,Segre char.\, & ~Possible cases~ & {\footnotesize D} & ~Constraints on $\beta_{n}$~\tabularnewline
\hline 
\hline 
\multirow{5}{*}{I} & \multirow{5}{*}{$[1111]$} & $[(\boldsymbol{1111})]$ &  & \tabularnewline
\cline{3-5} 
 &  & $[(\boldsymbol{11})(\boldsymbol{11})]$ &  & \tabularnewline
\cline{3-5} 
 &  & $[(\boldsymbol{111})1]$ & {*} & $\Delta=0$, $A\ne0$\tabularnewline
\cline{3-5} 
 &  & $[(\boldsymbol{11})11]$ & {*} & $\Delta=0$, $A=0$\tabularnewline
\cline{3-5} 
 &  & $[1111]$ & {*} & $\beta_{1}=\beta_{2}=\beta_{3}=0$\tabularnewline
\hline 
\multirow{4}{*}{IIa} & \multirow{4}{*}{$[211]$} & $[(\boldsymbol{211})]$ &  & $\Delta=0$\tabularnewline
\cline{3-5} 
 &  & $[(\boldsymbol{21})1]$ & {*} & $\Delta=0$, $A=0$\tabularnewline
\cline{3-5} 
 &  & $[2(\boldsymbol{11})]$ & {*} & $\Delta=0$, $A=0$\tabularnewline
\cline{3-5} 
 &  & $[211]$ & {*} & $\beta_{1}=\beta_{2}=\beta_{3}=0$\tabularnewline
\hline 
IIb & $[z\bar{z}11]$ & $\to$ Type I &  & \tabularnewline
\hline 
\multirow{2}{*}{III} & \multirow{2}{*}{$[31]$} & $[(\boldsymbol{31})]$ &  & $\Delta=0$\tabularnewline
\cline{3-5} 
 &  & $[31]$ &  & $\beta_{1}=\beta_{2}=\beta_{3}=0$\tabularnewline
\hline 
\end{tabular}\egroup\caption{\label{tab:segre}The Segre characteristics of all possible square
roots $S$. The highlighted eigenvalues are functions of the $\beta$-parameters.
The asterisk {*} in the {\footnotesize D}-column indicates a decoupled
case with some of the eigenvalues being arbitrary. The constraints
are given in terms of $\Delta\protect\coloneqq B^{2}-4AC$, $A\protect\coloneqq\beta_{2}^{2}-\beta_{1}\beta_{3}$,
$B\protect\coloneqq\beta_{1}\beta_{2}-\beta_{0}\beta_{3}$, and $C\protect\coloneqq\beta_{1}^{2}-\beta_{0}\beta_{2}$.}
\end{table}

\section{\label{sec:background}Background: Bimetric Field Equations}

The Hassan-Rosen (HR) action reads \cite{Hassan:2011zd},
\begin{align}
S_{\mathrm{HR}} & =\frac{1}{2}M_{g}^{d-2}\!\int\dd^{d}x\sqrt{-g}\,R_{g}+\frac{1}{2}M_{g}^{d-2}\!\int\dd^{d}x\sqrt{-f}\,R_{f}\nonumber \\
 & \qquad-\,m^{d}\!\int\dd^{d}x\sqrt{-g}\,V(S).\label{eq:hr-action}
\end{align}
It consists of two ordinary Einstein-Hilbert terms with Planck masses
$M_{g}$ and $M_{f}$, and the interaction term with the potential,\vspace{-0.9em}
\begin{equation}
V(S)\coloneqq\sum_{n=0}^{d}\beta_{n}\,e_{n}(S).\label{eq:V}
\end{equation}
Here, $S$ is the square root matrix function of the (1,1) tensor
field $g^{\mu\rho}f_{\rho\nu}$ (in matrix notation, $S=\sqrt{g^{-1}f}$).
The scalar invariant coefficients $e_{n}(S)$ in \eqref{eq:V} are
the elementary symmetric polynomials obtained through the generating
function \cite{macdonald:1998a},
\begin{equation}
E(t,S)=\det(I+tS)=\sum_{n=0}^{\infty}e_{n}(S)\,t^{n},\label{eq:e-genf}
\end{equation}
where $e_{n>d}(S)=0$ due to the Cayley-Hamilton theorem. In light
of the $E(t,S)$, the potential $V(S)$ can be seen as a linear combination
of the span of the $\beta$-parameters $\{\beta_{n}\}$, which are
the free parameters of the theory.

Variation of the action \eqref{eq:hr-action} yield two sets of equations
of motion in operator form \cite{Hassan:2014vja},\bSe\label{eq:hr-eom}
\begin{align}
G_{g}{}^{\mu}{}_{\nu}+\frac{m^{d}}{M_{g}^{d-2}}V_{g}{}^{\mu}{}_{\nu}(S) & =0,\label{eq:hr-eom-g}\\
G_{f}{}^{\mu}{}_{\nu}+\frac{m^{d}}{M_{f}^{d-2}}V_{f}{}^{\mu}{}_{\nu}(S) & =0,\label{eq:hr-eom-f}
\end{align}
\eSe where $G_{g}$ and $G_{f}$ denote the Einstein tensors of $g$
and $f$, respectively, and the stress-energy-like contributions $V_{g}$
and $V_{f}$ of the potential \eqref{eq:V} have the structure,\bSe
\begin{align}
V_{g}(S) & =\sum_{n=0}^{d-1}\beta_{n}\sum_{k=0}^{n}(-1)^{n+k}e_{k}(S)\,S^{n-k},\label{eq:Vg}\\
V_{f}(S) & =\sum_{n=0}^{d-1}\beta_{d-n}\sum_{k=0}^{n}(-1)^{n+k}e_{k}(S^{-1})\,S^{-n+k}.\label{eq:Vf}
\end{align}
\eSe

We say that $g$ is an Einstein solution iff it satisfies the Einstein
field equations,
\begin{equation}
G_{g}{}^{\mu}{}_{\nu}+\Lambda_{g}\delta_{\nu}^{\mu}=0,
\end{equation}
where $\Lambda_{g}=\mathrm{const}$; this implies $V_{g}{}^{\mu}{}_{\nu}(S)=m^{d}M_{g}^{2-d}\Lambda_{g}\delta_{\nu}^{\mu}$.
Similarly, we say that $f$ is an Einstein solution iff $G_{f}{}^{\mu}{}_{\nu}+\Lambda_{f}\delta_{\nu}^{\mu}=0$
for some $\Lambda_{f}=\mathrm{const}$, which implies $V_{f}{}^{\mu}{}_{\nu}(S)=m^{d}M_{f}^{2-d}\Lambda_{f}\delta_{\nu}^{\mu}$.
Notably, $V_{g}$ and $V_{f}$ are not independent as they obey the
algebraic identity \cite{Hassan:2014vja},
\begin{equation}
V_{g}(S)+\det(S)\,V_{f}(S)=V(S).\label{eq:V-identity}
\end{equation}

\section{Definitions\vspace{-0.3em}}

For a set of variables $x_{1},\dots,x_{d},$ define the symmetric
function,\vspace{-0.7em}
\begin{equation}
\left\langle x_{1},\dots,x_{d}\right\rangle _{k}^{n}\coloneqq\sum_{i=0}^{n}\beta_{i+k}\,e_{i}(x_{1},\dots,x_{d}).
\end{equation}
This function shifts the degree of homogeneity of the elementary symmetric
polynomials for $k$, and truncates the generating function \eqref{eq:e-genf}
at $n$. The subscript $k$ is accordingly called an \emph{offset}.

For a repeated single variable $x_{1}=...=x_{n}\eqqcolon\lambda$,
we introduce the following convention,
\begin{equation}
\left\langle \lambda\right\rangle _{k}^{n}\coloneqq\left\langle \lambda,\dots,\lambda\right\rangle _{k}^{n}=\sum_{i=0}^{n}\binom{n}{i}\beta_{i+k}\lambda^{i}.
\end{equation}
Also, if an argument of $\left\langle \cdot\right\rangle _{k}^{n}$
is a matrix $X$, the eigenvalues of $X$ are used instead.

In this notation, obviously $V(S)\coloneqq\left\langle S\right\rangle _{0}^{d}$,
and after some simplification, \eqref{eq:Vg} can be expressed,
\begin{equation}
V_{g}(S)=\sum_{n=0}^{d}(-1)^{n}\left\langle S\right\rangle _{n}^{d-n}S^{n}.
\end{equation}

\section{\label{sec:proof}Algebraic Forms of The Potential\vspace{-0.3em}}

In general, the square root matrix $S$ is not always diagonalizable;
yet it can always be put into the Jordan normal form. As shown in
\cite{Hassan:2017a}, the matrix $S$ will contain at most one Jordan
block up to size three, or at most one complex block, regardless of
spacetime dimension, which enables the algebraic classification of
bimetric solutions.

An Einstein solution in the $g$-sector implies $V_{g}(S)=\mathrm{const}.$
Any such constant can be absorbed in $\beta_{0}$ allowing one to
consider only the case $V_{g}(S)=0$. Including a minimal matter coupling
to one of the metrics does not introduce any further complication;
adding a stress-energy contribution on the left-hand side of \eqref{eq:hr-eom-g}
and setting $\tud{G_{g}}{\mu}{\nu}=M_{g}^{2-d}\tud{T_{g}}{\mu}{\nu}$,
again implies a constant $V_{g}$.

Since we are dealing with a system of algebraic equations, it is possible
to encounter degenerate cases for certain values of $\beta$-parameters
so that some of the eigenvalues of $S$ can be freely chosen (they
will not be functions of the $\beta$-parameters). Such cases will
be denoted as \emph{algebraically decoupled} and treated in \prettyref{sec:decoupled}.
Another categorization can be done according to the imposed conditions
on the $\beta$-parameters by the very algebraic structure of the
square root. If all $\beta$-parameters are independent of each other,
such a case will be called \emph{unconstrained}.

As we shall see, these two properties are expressible in terms of
the following auxiliary variables, for convenience defined here,\vspace{-0.2em}\bSe\label{eq:ABCD}
\begin{alignat}{2}
A & \coloneqq\beta_{2}^{2}-\beta_{1}\beta_{3}, & \quad B\coloneqq\, & \beta_{1}\beta_{2}-\beta_{0}\beta_{3},\label{eq:AB}\\
C & \coloneqq\beta_{1}^{2}-\beta_{0}\beta_{2}, & \Delta\coloneqq\, & B^{2}-4AC.\label{eq:CD}
\end{alignat}
\eSe

In the rest of this section, we explicitly solve this equation for
different types of the non-singular square roots matrix $S$ according
to the algebraic classification from \cite{Hassan:2017a} in four
dimensions. 

\subsection*{Type I\vspace{-0.5em}}

The square root $S$ of Type I has the diagonal form,
\begin{equation}
S_{\mathrm{I}}=\lambda_{1}\oplus\lambda_{2}\oplus\lambda_{3}\oplus\lambda_{4}=\diag(\lambda_{1},...,\lambda_{4}),\label{eq:tI-S}
\end{equation}
where $\lambda_{i}$ are real eigenvalues. Expanding the equation
$V_{g}(S_{\mathrm{I}})=0$ yields,\bSe
\begin{alignat}{2}
\left\langle \lambda_{2},\lambda_{3},\lambda_{4}\right\rangle _{0}^{3} & =0, & \left\langle \lambda_{1},\lambda_{3},\lambda_{4}\right\rangle _{0}^{3} & =0,\\
\left\langle \lambda_{1},\lambda_{2},\lambda_{4}\right\rangle _{0}^{3} & =0, & \quad\left\langle \lambda_{1},\lambda_{2},\lambda_{3}\right\rangle _{0}^{3} & =0.
\end{alignat}
\eSe These equations can also be seen as a homogeneous linear system
in $\beta_{0\le n\le3}$. Unless $\beta_{n}=0$, the discriminant
of the linear system must vanish identically,
\[
\begin{vmatrix}1 & \lambda_{2}+\lambda_{3}+\lambda_{4} & \lambda_{2}\lambda_{3}+\lambda_{4}\lambda_{3}+\lambda_{2}\lambda_{4} & \lambda_{2}\lambda_{3}\lambda_{4}\\
1 & \lambda_{1}+\lambda_{3}+\lambda_{4} & \lambda_{1}\lambda_{3}+\lambda_{4}\lambda_{3}+\lambda_{1}\lambda_{4} & \lambda_{1}\lambda_{3}\lambda_{4}\\
1 & \lambda_{1}+\lambda_{2}+\lambda_{4} & \lambda_{1}\lambda_{2}+\lambda_{4}\lambda_{2}+\lambda_{1}\lambda_{4} & \lambda_{1}\lambda_{2}\lambda_{4}\\
1 & \lambda_{1}+\lambda_{2}+\lambda_{3} & \lambda_{1}\lambda_{2}+\lambda_{3}\lambda_{2}+\lambda_{1}\lambda_{3} & \lambda_{1}\lambda_{2}\lambda_{3}
\end{vmatrix}=0,
\]
that is,\vspace{-1em}
\[
\prod_{1\le i<j\le4}(\lambda_{i}-\lambda_{j})=0.
\]
This means that at least two eigenvalues must be equal. Without loss
of generality, we can select $\lambda_{3}=\lambda_{4}\eqqcolon a$
(the permutation of the eigenvalues is a similarity transformation);
consequently,
\begin{align}
2\left\langle a\right\rangle _{0}^{2}+\left\langle a\right\rangle _{1}^{2}\left(\lambda_{1}+\lambda_{2}\right) & =0,\label{eq:tI-1}\\
\left\langle a\right\rangle _{1}^{2}\left(\lambda_{1}-\lambda_{2}\right) & =0,\label{eq:tI-2}\\
\left\langle \lambda_{1},\lambda_{2}\right\rangle _{0}^{2}+a\left\langle \lambda_{1},\lambda_{2}\right\rangle _{1}^{2} & =0.\label{eq:tI-3}
\end{align}
From \eqref{eq:tI-2}, we can have $\lambda_{1}=\lambda_{2}$ or $\left\langle a\right\rangle _{1}^{2}=0$.
Let us first consider the case when $\lambda_{1}=\lambda_{2}\eqqcolon b$,
which gives,
\begin{equation}
\left\langle a\right\rangle _{0}^{2}+b\left\langle a\right\rangle _{1}^{2}=0,\quad\left\langle b\right\rangle _{0}^{2}+a\left\langle b\right\rangle _{1}^{2}=0.\label{eq:tI-4}
\end{equation}
Adding and subtracting \eqref{eq:tI-4}, then redefining $a\equiv u+v$
and $b\equiv u-v$ in terms of two independent variables $u$ and
$v$, yields,
\begin{align}
\left\langle u\right\rangle _{0}^{3}-\beta_{2}v^{2}-\beta_{3}uv^{2} & =0,\label{eq:tI-6a}\\
v\left(\left\langle u\right\rangle _{1}^{2}-\beta_{3}v^{2}\right) & =0.\label{eq:tI-6b}
\end{align}
Let $v\ne0$, i.e., $\lambda_{3,4}=a\ne b=\lambda_{1,2}$, and $\left\langle u\right\rangle _{1}^{2}-\beta_{3}v^{2}=0$.
By using $\left\langle u\right\rangle _{0}^{3}=\left\langle u\right\rangle _{0}^{2}+u\left\langle u\right\rangle _{1}^{2}$,
\eqref{eq:tI-6a}\textendash \eqref{eq:tI-6b} become,
\begin{equation}
\left\langle u\right\rangle _{1}^{2}=\beta_{3}v^{2},\quad\ \left\langle u\right\rangle _{0}^{2}=\beta_{2}v^{2}.\label{eq:tI-7}
\end{equation}
The case $\beta_{2}=\beta_{3}=0$ implies $\beta_{0}=\beta_{1}=0$.
For $\beta_{2}\ne0$ or $\beta_{3}\ne0$, we get the solutions,
\begin{align}
u & =-B/(2A),\quad v=\pm\sqrt{\Delta}/(2A),\label{eq:tI-uv}
\end{align}
where the variables $A,B$, and the discriminant $\Delta$ are given
in \eqref{eq:ABCD}. Hence, \eqref{eq:tI-1}\textendash \eqref{eq:tI-3}
behaves as a single quadratic equation with the solutions $\lambda_{1}=\lambda_{2}=u\mp v$
and $\lambda_{3}=\lambda_{4}=u\pm v$.

On the other hand, let $v=0$ in \eqref{eq:tI-6a}\textendash \eqref{eq:tI-6b},
that is $a=b=u$. Then,
\begin{align}
\left\langle u\right\rangle _{0}^{3}=\beta_{0}+3\beta_{1}u+3\beta_{2}u^{2}+\beta_{3}u^{3} & =0.\label{eq:tI-cubic}
\end{align}
Depending on the $\beta$'s, \eqref{eq:tI-cubic} is at most cubic
in $u$ with up to three different solutions. For $\beta_{3}\ne0$,
the solutions to \eqref{eq:tI-cubic} are given by the Cardano formula,
\begin{align}
u_{1} & =\left[-\beta_{2}+(s+t)\right]/\beta_{3},\\
u_{2,3} & =\left[-\beta_{2}-{\textstyle \frac{1}{2}}(s+t)\pm\ii{\textstyle \frac{\sqrt{3}}{2}}(s-t)\right]/\beta_{3},
\end{align}
where,
\begin{gather}
t\equiv\left({\textstyle \frac{1}{2}}\beta_{3}B-\beta_{2}A+\sqrt{-\Delta_{3}}\right)^{1/3}\!,\quad s\equiv A/t,\\
\Delta_{3}\equiv A^{3}-\left({\textstyle \frac{1}{2}}\beta_{3}B-\beta_{2}A\right)^{2}.\quad
\end{gather}
Here, $\Delta_{3}$ is the discriminant of \eqref{eq:tI-cubic}; in
particular, \eqref{eq:tI-cubic} has three real roots for $\Delta_{3}\ge0$
(where at least two roots are equal for $\Delta_{3}=0$), otherwise
it has one real and two complex roots ($u_{1}$ is always real in
all cases).

Notably, the discriminant of \eqref{eq:tI-cubic} can be expressed
in terms of $\Delta$ from \eqref{eq:CD} as $\Delta_{3}=-\Delta\beta_{3}^{2}/4$.
This relation implies that, if all the cubic solutions $u$ of \eqref{eq:tI-cubic}
are real, then all $u\pm v$ from \eqref{eq:tI-uv} are complex. Also,
if one of the cubic solution is real and two complex conjugate, then
all $u\pm v$ from \eqref{eq:tI-uv} are real.

Now, we return to \eqref{eq:tI-1}\textendash \eqref{eq:tI-3} and
consider the case $\left\langle a\right\rangle _{1}^{2}=0$, which
implies $\left\langle a\right\rangle _{0}^{2}=0$ with the condition
$\Delta=0$. Then, if any of $\beta_{2}$, $\beta_{3}$ is vanishing,
all the eigenvalues are equal; otherwise, when $\beta_{2}\ne0$, $\beta_{3}\ne0$,
we have a decoupled case with an arbitrary $\lambda_{1}$ direct summed
with a block of three equal eigenvalues,
\begin{equation}
\lambda_{2}=a=(-\beta_{2}\pm\sqrt{A})/\beta_{3}=-B/(2A).
\end{equation}
where the last equality follows from $\Delta=B^{2}-4AC=0$ $\iff$
$A^{3}-\left({\textstyle \frac{1}{2}}\beta_{3}B-\beta_{2}A\right)^{2}=0$.
In the limit $A\to0$ we have $B\to0$, and $a=-B/(2A)\to\beta_{0}/\beta_{1}=\beta_{2}/\beta_{3}$,
rendering both $\lambda_{1}$ and $\lambda_{2}$ arbitrary.

Note that requiring $\Delta=0$ would constrain the $\beta$-parameters.
Another way to constrain the $\beta$-parameters is to impose a condition
on some of the eigenvalues in \eqref{eq:tI-cubic}. For example, requiring
$u=1$ to be one of the solutions imposes,
\begin{equation}
\beta_{0}+3\beta_{1}+3\beta_{2}+\beta_{3}=0,
\end{equation}
known as the asymptotic flatness condition for $f=g$ at infinity,
appearing, e.g., in \cite{Francesco:2017a}.

Concluding this section, we count the number of solutions in Type
I. In general, we have 3 branches in the cubic solution for $u$,
and 6 branches in different combinations of $a$ and $b$: $2\times aabb$,
$2\times abab$ and $2\times baab$. Note the symmetry in exchange
$a\leftrightarrow b$, so we do not calculate 6 branches in $a$ and
$b$ twice. This yields 9 solutions in total of which some are necessarily
complex.

The cubic solutions $\lambda_{1}=...=\lambda_{4}=u\ne0$, $v=0$ imply
proportional metrics, and at least one of the cubic solutions is real.
The quadratic solutions $\lambda_{1}=\lambda_{2}=u-v$, $\lambda_{3}=\lambda_{4}=u+v$
imply block proportional metrics. As noted, if all the cubic solutions
are real, then there are no quadratic solutions enabling the block
proportional metrics. As a consequence, we have either 3 real solutions
yielding proportional metrics, or 7 real solutions of which one is
giving proportional metrics and 6 giving different combinations of
block proportional metrics.

\subsection*{Type IIa\vspace{-0.5em}}

The square root of Type IIa has the block form,
\begin{equation}
S_{\mathrm{IIa}}=\begin{pmatrix}\lambda_{1} & 1\\
0 & \lambda_{1}
\end{pmatrix}\oplus\lambda_{2}\oplus\lambda_{3},
\end{equation}
where $\lambda_{i}$ are real eigenvalues. 

Expanding the equation $V_{g}(S_{\mathrm{IIa}})=0$ yields,\bSe
\begin{alignat}{2}
\left\langle \lambda_{1},\lambda_{2},\lambda_{3}\right\rangle _{0}^{3} & =0, & \left\langle \lambda_{2},\lambda_{3}\right\rangle _{1}^{2} & =0,\label{eq:tIIa-1}\\
\left\langle \lambda_{1},\lambda_{1},\lambda_{3}\right\rangle _{0}^{3} & =0, & \quad\left\langle \lambda_{1},\lambda_{1},\lambda_{2}\right\rangle _{0}^{3} & =0.\label{eq:tIIa-2}
\end{alignat}
\eSe The first equation in \eqref{eq:tIIa-1} can be expanded $\left\langle \lambda_{2},\lambda_{3}\right\rangle _{0}^{2}+\lambda_{1}\left\langle \lambda_{2},\lambda_{3}\right\rangle _{1}^{2}=0$,
thus $\left\langle \lambda_{2},\lambda_{3}\right\rangle _{0}^{2}=0$.
Also, the sum and the difference of the equations in \eqref{eq:tIIa-2},
gives,
\begin{align}
\left\langle \lambda_{2},\lambda_{3}\right\rangle _{0}^{2}=\left\langle \lambda_{2},\lambda_{3}\right\rangle _{1}^{2} & =0,\\
2\left\langle \lambda_{1}\right\rangle _{0}^{2}+\left\langle \lambda_{1}\right\rangle _{1}^{2}\left(\lambda_{3}+\lambda_{2}\right) & =0,\\
\left\langle \lambda_{1}\right\rangle _{1}^{2}\left(\lambda_{3}-\lambda_{2}\right) & =0.
\end{align}
Consider the case $\lambda_{2}=\lambda_{3}\eqqcolon a$. Then,
\begin{equation}
\left\langle a\right\rangle _{0}^{2}=\left\langle a\right\rangle _{1}^{2}=0.\label{eq:tIIa-5}
\end{equation}
For $\beta_{2}=\beta_{3}=0$ we have $\beta_{0}=\beta_{1}=0$. For
$\beta_{2}=0$, $\beta_{3}\ne0$ we get $\beta_{0}^{2}\beta_{3}+4\beta_{1}^{3}=0$
and the solution $\lambda_{1}=a=-\beta_{0}/(2\beta_{1})$. For $\beta_{3}=0$,
$\beta_{2}\ne0$, we get $4\beta_{0}\beta_{2}=3\beta_{1}^{2}$ and
the solution $\lambda_{1}=a=-\beta_{1}/(2\beta_{2})$. Finally, for
$\beta_{2}\ne0$, $\beta_{3}\ne0$, we get $\Delta=0$, and the solutions,
\begin{equation}
\lambda_{1}=a=(-\beta_{2}\pm\sqrt{A})/\beta_{3}=-B/(2A).\label{eq:tIIa-7}
\end{equation}
A similar analysis as in Type I, gives a decoupled case with $a=-\beta_{2}/\beta_{3}$
and an arbitrary $\lambda_{1}$ in the limit $A\to0.$

Now, consider the case $\left\langle \lambda_{1}\right\rangle _{1}^{2}=0$.
This implies $\left\langle \lambda_{1}\right\rangle _{0}^{2}=\left\langle \lambda_{1}\right\rangle _{1}^{2}=0$,
which produces the same constraints as in the case above for $\lambda_{2}=\lambda_{3}=a$
but with $\lambda_{1}$ in place of $a$ in \eqref{eq:tIIa-5}. On
the other hand, defining $\lambda_{2}\equiv u+v$, $\lambda_{3}\equiv u-v$
again yields the same $u$ and $v$ as in \eqref{eq:tI-uv}. Substituting
the constraint on $\beta_{0}$ from $\left\langle \lambda_{1}\right\rangle _{0}^{2}=\left\langle \lambda_{1}\right\rangle _{1}^{2}=0$
into $u$ and $v$ yields $v=0$, and $\lambda_{2}=\lambda_{3}=u$.
Hence we again get \eqref{eq:tIIa-7}. 

Therefore, we always have $\lambda_{1}=\lambda_{2}=\lambda_{3}$ in
Type IIa, with the constraint $\Delta=0$ on the $\beta$-parameters.

\subsection*{Type IIb\vspace{-0.5em}}

The square root of Type IIb has the block form,
\begin{equation}
S_{\mathrm{IIb}}=\begin{pmatrix}a & -b\\
b & a
\end{pmatrix}\oplus\lambda_{1}\oplus\lambda_{2},\label{eq:tIIb-S}
\end{equation}
where $\lambda_{1}$ and $\lambda_{2}$ are two real eigenvalues,
while $a$ and $b$ are two real numbers representing a complex conjugate
eigenvalue pair $a\pm\mathrm{i}\,b$. In the limit $a\to0$, one gets
Type IV as a branch-cut, while $b\to0$ produces Type I.

The equation $V_{g}(S_{\mathrm{IIb}})=0$ yields,\bSe
\begin{align}
\left\langle \lambda_{1},\lambda_{2}\right\rangle _{0}^{2}+a\left\langle \lambda_{1},\lambda_{2}\right\rangle _{1}^{2} & =0,\label{eq:tIIb-1}\\
b\left\langle \lambda_{1},\lambda_{2}\right\rangle _{1}^{2} & =0.\label{eq:tIIb-2}\\
\left\langle a\right\rangle _{0}^{2}+\beta_{2}b^{2}+\left(\left\langle a\right\rangle _{1}^{2}+\beta_{3}b^{2}\right)\lambda_{2} & =0,\label{eq:tIIb-3}\\
\left\langle a\right\rangle _{0}^{2}+\beta_{2}b^{2}+\left(\left\langle a\right\rangle _{1}^{2}+\beta_{3}b^{2}\right)\lambda_{1} & =0.\label{eq:tIIb-4}
\end{align}
\eSe Multiplying \eqref{eq:tIIb-2} by $a/b$ and subtracting from
\eqref{eq:tIIb-1}, then adding and subtracting \eqref{eq:tIIb-3}
and \eqref{eq:tIIb-4} gives,
\begin{align}
\left\langle \lambda_{1},\lambda_{2}\right\rangle _{0}^{2}=\left\langle \lambda_{1},\lambda_{2}\right\rangle _{1}^{2} & =0,\label{eq:tIIb-6}\\
2\left(\left\langle a\right\rangle _{0}^{2}+\beta_{2}b^{2}\right)+\left(\left\langle a\right\rangle _{1}^{2}+\beta_{3}b^{2}\right)\left(\lambda_{1}+\lambda_{2}\right) & =0,\label{eq:tIIb-5}\\
\left(\left\langle a\right\rangle _{1}^{2}+\beta_{3}b^{2}\right)\left(\lambda_{1}-\lambda_{2}\right) & =0.
\end{align}
For $\lambda_{1}=\lambda_{2}\eqqcolon c$ we have,\vspace{-0.3em}
\begin{align}
\left\langle c\right\rangle _{0}^{2}=\left\langle c\right\rangle _{1}^{2} & =0,\\
\left\langle a\right\rangle _{0}^{2}+\beta_{2}b^{2}+\left(\left\langle a\right\rangle _{1}^{2}+\beta_{3}b^{2}\right)c & =0.
\end{align}
As earlier, the first equation imposes a condition on the $\beta$-parameters,
fixing $c$. Setting $a=u+c$, gives $(u^{2}+b^{2})(\beta_{2}+c\beta_{3})=0$.
The case $\beta_{2}+c\beta_{3}=0$ is equivalent to $\beta_{2}^{2}-\beta_{1}\beta_{3}=0$,
yielding arbitrary $a$ and $b$. Otherwise, $\beta_{2}+c\beta_{3}\ne0$
implies $a+\ii\,b=c$. This is only possible if $a=c$ and $b=0$,
which is then a subset of Type I.

For $\lambda_{1}\ne\lambda_{2}$ and $\left\langle a\right\rangle _{1}^{2}+\beta_{3}b^{2}=0$,
the solutions $\lambda_{1}$ and $\lambda_{2}$ are obtained from
\eqref{eq:tIIb-6}, so that $\lambda_{1}=u+v$ and $\lambda_{2}=u-v$
with $u$ and $v$ again given in \eqref{eq:tI-uv}. On the other
hand, we have $\left\langle a\right\rangle _{0}^{2}+\beta_{2}b^{2}=0$
and $b^{2}=-\beta_{3}/\left\langle a\right\rangle _{1}^{2}=-\beta_{2}/\left\langle a\right\rangle _{0}^{2}$,
implying $\left\langle a\right\rangle _{0}^{2}\beta_{3}=\beta_{2}\left\langle a\right\rangle _{1}^{2}$
and,
\begin{equation}
a=-B/(2A)=u,\quad b^{2}=-\left\langle a\right\rangle _{0}^{2}/\beta_{2}=-v^{2}.
\end{equation}
This shows that either $b$ or $v$ must be imaginary, unless $b=0$,
which is a subset of to Type I. This is not surprising as Type IIb
can be expressed as Type I having a pair of complex conjugate eigenvalues.
Since the Lorentzian signature of the metrics forbids the existence
of two complex blocks, Type I cannot be block proportional with two
pairs of complex numbers. Therefore, there are no solutions of Type
IIb which are different than of Type I.

\subsection*{Type III\vspace{-0.5em}}

The square root of Type III has the form,
\begin{equation}
S_{\mathrm{III}}=\begin{pmatrix}\lambda_{1} & 1 & 0\\
0 & \lambda_{1} & 1\\
0 & 0 & \lambda_{1}
\end{pmatrix}\oplus\lambda_{2},
\end{equation}
where $\lambda_{1}$ and $\lambda_{2}$ are two real eigenvalues.
The equation $V_{g}(S_{\mathrm{III}})=0$ yields,\bSe
\begin{alignat}{2}
\left\langle \lambda_{1},\lambda_{1},\lambda_{2}\right\rangle _{0}^{3} & =0, & \quad\left\langle \lambda_{2}\right\rangle _{2}^{1} & =0,\\
\left\langle \lambda_{1},\lambda_{2}\right\rangle _{1}^{2} & =0, & \left\langle \lambda_{1}\right\rangle _{0}^{3} & =0.
\end{alignat}
\eSe Expanding $\left\langle \lambda_{1},\lambda_{1},\lambda_{2}\right\rangle _{0}^{2}=\left\langle \lambda_{1},\lambda_{2}\right\rangle _{0}^{2}+\lambda_{1}\left\langle \lambda_{1},\lambda_{2}\right\rangle _{1}^{2}$
gives $\left\langle \lambda_{1},\lambda_{2}\right\rangle _{0}^{2}=0$
that together with $\left\langle \lambda_{1},\lambda_{2}\right\rangle _{1}^{2}=0$
completely determines $\lambda_{1}$ and $\lambda_{2}$ so that $\lambda_{1}=u+v$
and $\lambda_{2}=u-v$ with $u$ and $v$ given in \eqref{eq:tI-uv}.
On the other hand, $\left\langle \lambda_{2}\right\rangle _{2}^{1}=0$
is a linear equation in $\lambda_{2}$ with the solution $\lambda_{2}=-\beta_{2}/\beta_{3}$.
Substituting $\left\langle \lambda_{1},\lambda_{2}\right\rangle _{1}^{2}=0$
gives $A=0$, and further $\left\langle \lambda_{1},\lambda_{2}\right\rangle _{0}^{2}=0$
yields $\beta_{0}\beta_{3}^{2}=\beta_{2}^{3}$. Finally, $\left\langle \lambda_{1}\right\rangle _{0}^{3}=0$
gives $\lambda_{1}=\lambda_{2}$ so that both $v=0$ and the discriminant
of the cubic equation vanishes. In conclusion, we have only a single-eigenvalue
solution $\lambda_{1}=\lambda_{2}=-\beta_{2}/\beta_{3}$ with the
constraint $\beta_{0}/\beta_{1}=\beta_{1}/\beta_{2}=\beta_{2}/\beta_{3}$.

\subsection*{Type IV\vspace{-0.5em}}

This is the case of a non-primary square root that has the block form
as in Type IIb \eqref{eq:tIIb-S}, but with $a=0$ so that the complex
block represents a pair of imaginary eigenvalues $\pm\mathrm{i}\,b$.
The solutions are $-b^{2}=\lambda_{1}^{2}=\lambda_{2}^{2}$, which
cannot be satisfied for any non-vanishing real $b$ and $\lambda_{1}$
(unless $\Delta=A=0$).

\section{Algebraically Decoupled cases\label{sec:decoupled}\vspace{-0.3em}}

In particular for $\beta_{1}=\beta_{2}=\beta_{3}=0$, the kinetic
terms in \prettyref{eq:hr-action} decouple, so there is no algebraic
restriction on $S$ imposed by $V_{g}(S)=0$ (dynamical restrictions
can, however, come out from the field equations). A similar degeneracy
occurs when the $\beta$-parameters form a geometric progression,
$\beta_{k+1}/\beta_{k}=\mathrm{const}$.

During the enumeration of all possible solutions of the equation $V_{g}=0$,
we have encountered three possible decoupled cases:\vspace{-0.3em}
\begin{enumerate}[leftmargin=35mm,label=({\footnotesize D}\arabic*)]
\item \label{enu:D1}$\ \beta_{1}=\beta_{2}=\beta_{3}=0$,\vspace{-0.3em}
\item \label{enu:D2}$\ \Delta=0$, $A=0$,\vspace{-0.3em}
\item \label{enu:D3}$\ \Delta=0,$ $A\ne0$,
\end{enumerate}
where \ref{enu:D2} and \ref{enu:D3} only occur for non-vanishing
$\beta$-parameters. The condition \ref{enu:D2} states a geometric
progression $\beta_{k+1}/\beta_{k}=\mathrm{const}$.

For Type I and Type IIb, the conditions \ref{enu:D1}, \ref{enu:D2}
or \ref{enu:D3} render four, three or two arbitrary eigenvalues,
respectively. 

For Type IIa, the conditions \ref{enu:D1} or \ref{enu:D2} make two
or one of the eigenvalues arbitrary, respectively. However, \ref{enu:D3}
is mandatory for Type IIa; it must be satisfied for Type IIa to exist,
in which case all the eigenvalues are given by the $\beta$-parameters.

For Type III the eigenvalues are determined by the $\beta$-parameters.
Moreover, the condition \ref{enu:D2} must be satisfied for Type III
to exist.

\section{An Example Solution\label{sec:example}\vspace{-0.3em}}

In the analysis above, we assessed the algebraic restrictions on the
bimetric field equations. As a dynamical example, let us consider
a spherically symmetric Einstein solution in one sector, let's say
$g$.

By Birkhoff's theorem, any spherically symmetric solution is locally
isometric to a subset of the Schwarzschild solution. Therefore, without
loss of generality, we can consider the line element of $g$ in the
standard Schwarzschild chart $(t,r,\theta,\phi)$,
\[
\dd s_{g}^{2}=-F\dd t^{2}+F^{-1}\dd r^{2}+r^{2}\dd\theta^{2}+r^{2}\sin^{2}\theta\,\dd\phi^{2},
\]
where $F\equiv1-r_{\eH}/r$. Including a cosmological constant $\Lambda$,
we have $F=1-r_{\mathrm{H}}/r-\Lambda r^{2}/3$, yielding $G_{g}{}^{\mu}{}_{\nu}+\Lambda\delta_{\nu}^{\mu}=0$.
In the limit $r_{\eH}\to0$, $\Lambda\to0$, we recover the Minkowski
solution. Since the metric is diagonal, the square root is of Type
I as in \eqref{eq:tI-S} (if $g$ was in the Eddington-Finkelstein
chart, the square root could be of Type IIa or IIb). This implies,
\[
\dd s_{f}^{2}=-\lambda_{1}^{2}F\dd t^{2}+\lambda_{2}^{2}F^{-1}\dd r^{2}+\lambda_{3}^{2}r^{2}\dd\theta^{2}+\lambda_{4}^{2}r^{2}\sin^{2}\theta\,\dd\phi^{2},
\]
resulting in the Einstein operator for $f$,
\begin{equation}
G_{f}{}^{\mu}{}_{\nu}=\left[-\Lambda\lambda_{2}^{-2}+r^{2}\left(\lambda_{2}^{-2}-\lambda_{3}^{-2}\right)\right]\delta_{\nu}^{\mu}.\label{eq:ex-f-sector}
\end{equation}
Hence, the \emph{dynamics} imposes $\lambda_{2}^{2}=\lambda_{3}^{2}$
for $f$ to satisfy the Einstein field equations. This constraint
is absent for the Minkowski solution; nevertheless, in general, for
a more complicated $g$, constraining $S$ this way might not always
be possible.

On the other hand, \emph{algebraically}, the proved proposition states
that either $\lambda_{1}=\lambda_{2}$ and $\lambda_{3}=\lambda_{4}$,
or $\lambda_{1}=\lambda_{3}$ and $\lambda_{2}=\lambda_{4}$, or $\lambda_{1}=\lambda_{4}$
and $\lambda_{2}=\lambda_{3}$, or that all eigenvalues are equal.
Therefore, (depending on the $\beta$-parameters),
\[
\dd s_{f}^{2}=-\lambda_{1}^{2}F\dd t^{2}+\lambda_{2}^{2}F^{-1}\dd r^{2}+\lambda_{2}^{2}r^{2}\dd\theta^{2}+\lambda_{1}^{2}r^{2}\sin^{2}\theta\,\dd\phi^{2},
\]
is a healthy Einstein solution in the $f$-sector. This solution may
or may not be spherically symmetric.

After the chart transition $t^{\prime}=\lambda_{1}t$, $r^{\prime}=\lambda_{2}r$,
$\theta^{\prime}=\lambda_{2}\theta$, $\phi^{\prime}=\lambda_{1}\phi$,
one can show that $\lambda_{2}=1$ is needed for $f$ to be spherically
symmetric. This condition (put \emph{by hand}) leaves an arbitrary
$\lambda_{1}\eqqcolon c$. Then, in the setup with $S=\diag(c,1,1,c)$,
there are two sets of Killing vector fields (KVF) generating SO(3).
In particular, in the original chart $(t,r,\theta,\phi)$, the KVFs
that generate a \emph{separate} spherical symmetry for $f$ are,
\begin{align}
\eta_{1} & =c^{-1}\partial_{\phi},\\
\eta_{2} & =\cos(c\phi)\,\partial_{\theta}-\cot(\theta)\sin(c\phi)\,c^{-1}\partial_{\phi},\\
\eta_{3} & =-\sin(c\phi)\,\partial_{\theta}-\cot(\theta)\cos(c\phi)\,c^{-1}\partial_{\phi}.
\end{align}
In comparison, the KVFs of $g$ have a similar structure, but without
the presence of $c$. Finally, we take into account the \emph{topology}
of the Schwarzschild solution, which is diffeomorphic to $\mathbb{R}^{2}\times\mathcal{S}^{2}$.
This further imposes $c=\pm1$, which can be shown by considering
all possible scalars created from the KVFs of $g$ and $f$, for example,
$g(\xi_{2},\eta_{2})$. Namely, all of these are the invariant scalar
fields in $\phi$ involving the mixings of the trigonometric functions
of $\phi$ and $c\phi$. The $\mathcal{S}^{2}$ subspace topology
in an atlas adapted to $g$ requires the scalars to have the same
value at $\phi=0$ and $\phi=2\pi$, which constrains $c$ to be a
non-vanishing integer (a winding number of the orbits of $f$'s KVFs).
On the other hand, we can apply the same argument imposing the same
topology in an atlas adapted to $f$ (now in terms of some azimuth
$\psi$). This requires $1/c$ to be a non-vanishing integer, too.
Thus, necessarily $c^{2}=1$.

As just illustrated, bi-Einstein solutions may not always exist, even
when in vacuum and with the algebraic restrictions satisfied (because
the square root must satisfy additional constraints imposed by dynamics,
which might not always be possible). Moreover, the symmetries in the
two sectors may or may not be related, and the topology must also
be taken in account. Apart from the symmetries and the topology, there
are also remaining questions about what are the additional (general)
constraints implied by the dynamics of bi-Einstein field equations
(with or without matters sources). All these questions will be addressed
elsewhere in more detail.
\begin{acknowledgments}
We are grateful to Ingemar Bengtsson, Bo Sundborg and Anders Lundkvist
for helpful discussions.
\end{acknowledgments}

\end{document}